# ON SOME PROCESSES AND DISTRIBUTIONS IN A COLLECTIVE MODEL OF INVESTORS' BEHAVIOR

Kyrylo Shmatov, Mikhail Smirnov

**Abstract**

This article considers a model for alternative processes for securities prices and compares this model with actual return data of several securities. The distributions of returns that appear in the model can be Gaussian as well as non–Gaussian; in particular they may have two peaks. We consider a discrete Markov chain model. This model in some aspects is similar to well–known Ising model describing ferromagnetics. Namely we consider a set of $N$ investors, each of whom has either bullish or bearish opinion, denoted by plus or minus respectively. At every time step each of $N$ investors can change his/her sign. Let denote by $p_+(n)$ the probability of a plus becoming a minus and denote by $p_-(n)$ a probability of a minus becoming a plus, assuming that probabilities depend only on the bullish sentiment described as the number $n$ of bullish investors among the total of $N$ investors. The number of bullish investors $n(t)$ forms a Markov chain whose transition matrix is calculated explicitly. The transition matrix of that chain is ergodic and any initial distribution of bullish investors converges to stationary. Stationary distributions of bullish investors in Markov chain model for $p_+(n) = c\exp(-(k(\frac{n}{N} - \frac{1}{2}) + h))$, $p_-(n) = c\exp(k(\frac{n}{N} - \frac{1}{2}) + h)$ are similar to continuous distributions of the "theory of social imitation" of Callen and Shapero. Distributions obtained this way can represent 3 types of market behavior: one peaked-distribution that is close to Gaussian, transition market (flattening of the top), and two-peaked distribution.

## 1. Introduction

This paper presents an agent based approach to stock market modeling in many aspects similar to the well–known Ising model describing ferromagnetics. Historically, E. Callen and D. Shapero [1] proposed that model similar to Ising might be extrapolated to the theory of organized social behavior and developed a nonlinear statistical model that they called "theory of social imitation." According to this theory the polarization of political opinion in social groups undergoes transitions between states of macroscopic disorder and more orderly coherent states as well as it does in a ferromagnetic. T. Vaga [2] then applied these ideas to stock markets. According to Vaga's model markets do not always exhibit random walk properties. Random–walk periods are followed by periods of choppy or trending (crowd) behavior. Transitions from random–walk markets to periods of crowd behavior are characterized by instability. Crowd behavior can lead to a choppy market or to the most rewarding investment opportunities, trending markets that should be of special interest to investors. Crowd behavior is defined here as a state of order in a complex system comprising a large number of independently acting subsystems.

Recently a number of authors working on market dynamics and agent based approaches obtained important results. Especially worth mentioning are works of Kaizoji [3], Kaizoji, Bornholdt and Fujiwara [4] and Krawiecki and Holyst, [5]. Some of the results i.e. Challet, Marsili, Martino [6] are in the area in econophysics literature called *minority games,* the situations where agents strive to belong to the global minority.
The other group o results is based on the voter model where agents strive to belong to majority and that result in herding behavior Chamley [7], Granovsky, Madras [8].

recently there were results on models combining minority and majority features Kaizoji, Bornholdt, Fujivara [4], Badshah, Boyer and Theodosopoulos [9], [10]. We refer to works [11-21] for different aspects of agent-based models.

We study a model based on certain general assumptions about the group of investors. Under some "normal" conditions individuals would act independently of each other. Under some specific conditions, however, the same individuals may begin thinking polarized, i.e., the individuals will act as a crowd and individual rational thinking will be replaced by a collective "group think." Transitions from disorder to order (and vice versa) tend to share the same macroscopic characteristics. Investor opinions and the prevailing bias in economic fundamentals control the state of the market. When investor opinion is not conducive to crowd behavior, the market is likely to exhibit random walk properties (efficient market). The combination of strong positive fundamentals and investor sentiment conducive to crowd behavior leads to the safest and most rewarding state, a trending bull market. In these market periods, a fully invested position is necessary to avoid the risk of underperforming the market averages. When the fundamental bias is neither strongly bullish nor bearish during a period of crowd behavior, it results in a dangerously volatile "double-peaked" market state where crowd opinion may switch abruptly from bullish to bearish and vice versa. In addition to these most prevalent market states, a trending bear market state exists which exhibits highly negative fundamentals with crowd behavior. When crowd behavior prevails, market fluctuations will tend to follow a bimodal distribution.

## 2. Markov Chain Approach

Consider a set of $N$ investors, each of whom has either bullish or bearish opinion, denoted by plus or minus respectively. At each unit time step investor may change his opinion. Let $p_+(n)$ denote the probability of a plus becoming a minus and $p_-(n)$ probability of a minus become a plus, where $n$ is the number of plus signs among total $N$ investors. Define also complementary probabilities

$q_+(n) = 1 - p_+(n)$,
$q_-(n) = 1 - p_-(n)$.

We want to study the evolution for $t=0, 1, 2, 3,\ldots$ of the number $n(t)$ of plus signs among $N$ investors. That number $n(t)$ form a Markov chain.

We start with some initial probability distribution of number of bullish investors (pluses) at time 0. Let $f(n,0)$ be the probability that at time $t=0$ the number of bullish investors is exactly $n$, $n=0\ldots N$. So initial probability distribution is a vector of $N+1$ non–negative numbers $f(n,0)$ with sum 1 (because the total probability is 1). For example the initial probability distribution having there are exactly $m$ bullish investors with probability 1 is given by $f(n,0)=\delta(n-m)$, where $\delta(n-m)=0$ when $n \neq m$, and $\delta(n-m)=1$ when $n=m$.

Now let $f(n,t)$ be the probability that at time $t$ the number of bullish investors is exactly $n$. The time evolution of the column of probability distribution given by $f(n,t)$ can be described by

$$\begin{bmatrix} f(0,t+1) \\ f(1,t+1) \\ \ldots \\ f(N,t+1) \end{bmatrix} = W_{(N+1)\times(N+1)} \begin{bmatrix} f(0,t) \\ f(1,t) \\ \ldots \\ f(N,t) \end{bmatrix}$$

Components of the conditional $(N+1)\mathsf{x}(N+1)$ probability matrix $W$ can be found by explicit counting of probabilities of all transitions from all counts of bullish investors at step $t$ leading to exactly $n$ bullish investors at step $t+1$.

The other representation of a matrix element in $m$–th row and $n$–th column $w_{mn}$ is that it is equal to the coefficient near $y^m$ of a polynomial

$$Q_n(y) = (p_-(n)y + q_-(n))^{N-n}(q_+(n)y + p_+(n))^n.$$

That coefficient is equal to

$$w_{mn} = \frac{1}{m!}\frac{\partial^m}{\partial y^m}\left((p_-(n)y + q_-(n))^{N-n}(q_+(n)y + p_+(n))^n\right)\Big|_{y=0}$$

and in explicit form is

$$w_{mn} = \sum_{j=\max\{0,m-n\}}^{\min\{m,N-n\}} \binom{N-n}{j}\binom{n}{m-j} p_-^j(n) q_-^{N-n-j}(n) q_+^{m-j}(n) p_+^{n-m+j}(n).$$

So

$$f(m,t+1) = \sum_{n=0}^{N} \frac{1}{m!}\frac{\partial^m}{\partial y^m}\left((p_-(n)y + q_-(n))^{N-n}(q_+(n)y + p_+(n))^n\right) f(n,t)\Big|_{y=0}$$

For example, in the case of $N=3$ matrix $W$ has the form

$$\begin{pmatrix} q_-^3(0) & q_-^2(1)p_+(1) & q_-(2)p_+^2(2) & p_+^3(3) \\ 3p_-(0)q_-^2(0) & q_-^2(1)q_+(1)+2p_-(1)q_-(1)p_+(1) & 2q_-(2)q_+(2)p_+(2)+p_-(2)p_+^2(2) & 3q_+(3)p_+^2(3) \\ 3p_-^2(0)q_-(0) & 2p_-(1)q_-(1)q_+(1)+p_-^2(1)p_+(1) & q_-(2)q_+^2(2)+2p_-(2)q_+(2)p_+(2) & 3q_+^2(3)p_+(3) \\ p_-^3(0) & p_-^2(1)q_+(1) & p_-(2)q_+^2(2) & q_+^3(3) \end{pmatrix}.$$

One can explicitly show that the sum of numbers in each column equals 1. Indeed by our formula elements of the $n$–th column are coefficients of $Q_n(y)$ and sum of elements in the $n$–th column is just the sum of coefficients of that polynomial and that is

$$Q_n(1) = (p_-(n)1 + q_-(n))^{N-n}(q_+(n)1 + p_+(n))^n = 1^{N-n}1^n = 1.$$

The matrix $W$ is ergodic, because it is time–independent and $w_{ij} \geq \delta$, where $\delta$ is independent of $i, j$ and time (one cane take the smallest value of $w_{ij}$ for $\delta$). That means any initial distribution $f(n,0)$ at time zero goes to a stationary distribution as time increases. Eigenvectors of $W$ corresponding to an eigenvalue of 1 deliver stationary probability distribution $f(n, +\infty)$ for given functions $p_+(n)$, $p_-(n)$. For general properties of Markov chains see for example Rozanov [22].

Now we can formulate different versions of this Markov chain model.
We are given two sets of $N$ numbers

$$p_+(1), p_+(2), p_+(3),\ldots, p_+(N),$$
$$p_-(0), p_-(1), p_-(3),\ldots, p_-(N-1),$$

note that $p_+(0)$ and $p_-(N)$ are not used in matrix $W$. We want to study qualitative properties of a stationary distribution given by the Markov chain with transition matrix $W$ constructed from those $p_+(n)$ and $p_-(n)$. Probabilities $p_+(n)$ and $p_-(n)$ can have different functional form thus leading to different versions of the model.

### *"Ising" model* (Discrete version of "Social Imitation Model")

$$p_+(n) = c\exp\left(-\left(k\left(\tfrac{n}{N}-\tfrac{1}{2}\right)+h\right)\right),$$
$$p_-(n) = c\exp\left(k\left(\tfrac{n}{N}-\tfrac{1}{2}\right)+h\right)$$

$k$ is any non–negative number, $h$ is any real number, $c$ is normalizing constant.

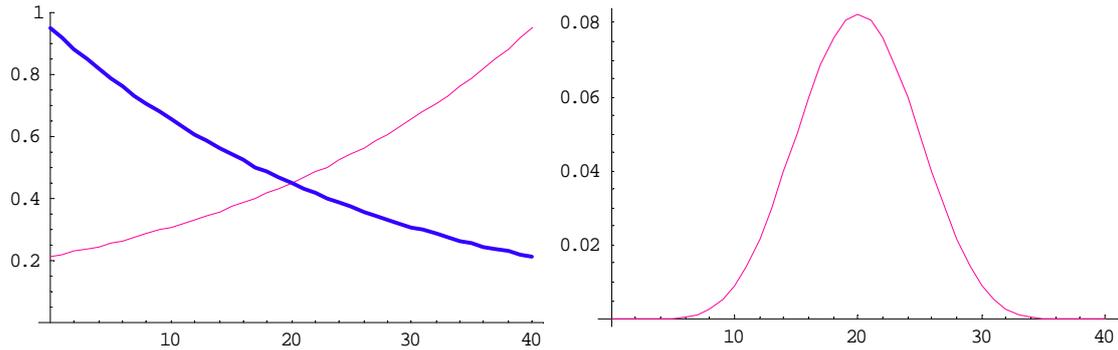

Fig 1. SI Model. On the left $p_+(n)$ thick line, $p_-(n)$, on the right probability density function after 300 iterations; number of plus investors $n$ is plotted on the x-axis. $N=40$, $k=1.5$; $h=0$; $c=0.45$.

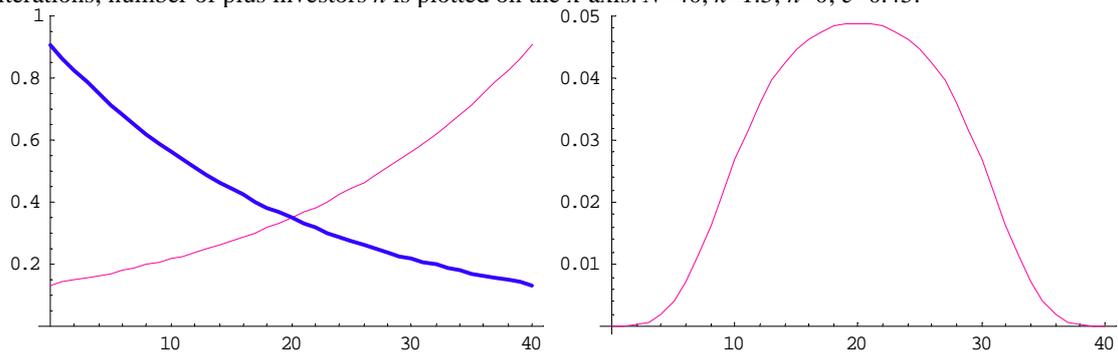

Fig 2. SI Model. On the left $p_+(n)$ thick line, $p_-(n)$, on the right PDF, $n$ is plotted on the x-axis. $N=40$, $k=1.9$; $h=0$; $c=0.35$.

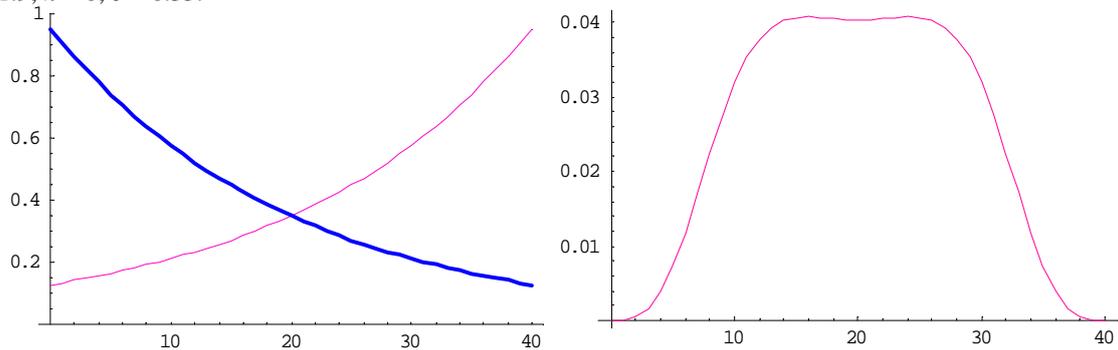

Fig 3. SI Model. On the left $p_+(n)$ thick line, $p_-(n)$, on the right PDF, $n$ is plotted on the x-axis. $N=40$, $k=2$; $h=0$; $c=0.35$.

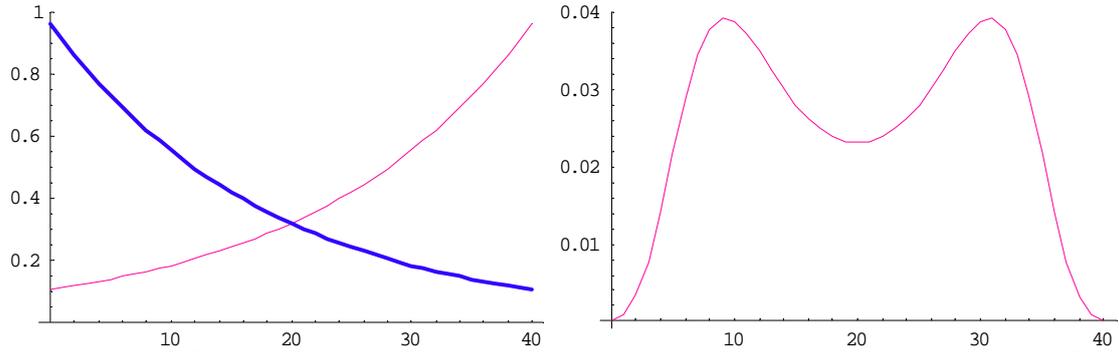
Fig 4. SI Model. On the left $p_+(n)$ thick line, $p_-(n)$, on the right PDF, $n$ is plotted on the x-axis. $N=40$, $k=2.2$; $h=0$; $c=0.32$;

### Chen's socio–psychological model of collective choice (Chen [23-24])

$$n\, p_+(n) = a_1 n + b_1 n (N-n)$$
$$(N-n)\, p_-(n) = a_2 (N-n) + b_2 n (N-n)$$

This transition probability has simple explanation. The rates of changes in bullish opinion $n\, p_+(n)$ and bearish opinion $(N-n)\, p_-(n)$ depend both on the population size holding the same opinion and the interaction between the people holding opposite opinions. We can rewrite $p_+(n)$, $p_-(n)$ as

$$p_+(n) = a_1 + b_1 (N-n)$$
$$p_-(n) = a_2 + b_2 n$$

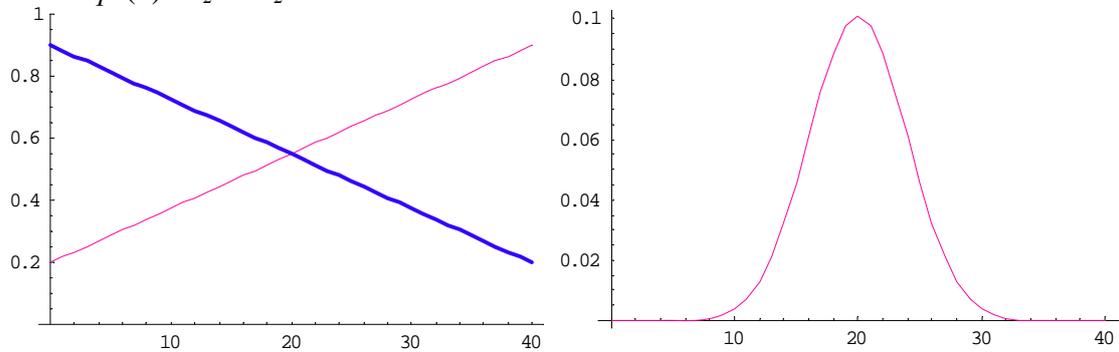
Fig 5. Chen's model. Left $p_+(n)$, $p_-(n)$, right probability density, $n$ is plotted on the x-axis. $N=40$, $a_1 = a_2 = 0.2$; $b_1 = b_2 = 0.7/40 = 0.0175$; $b_1 < a_1$

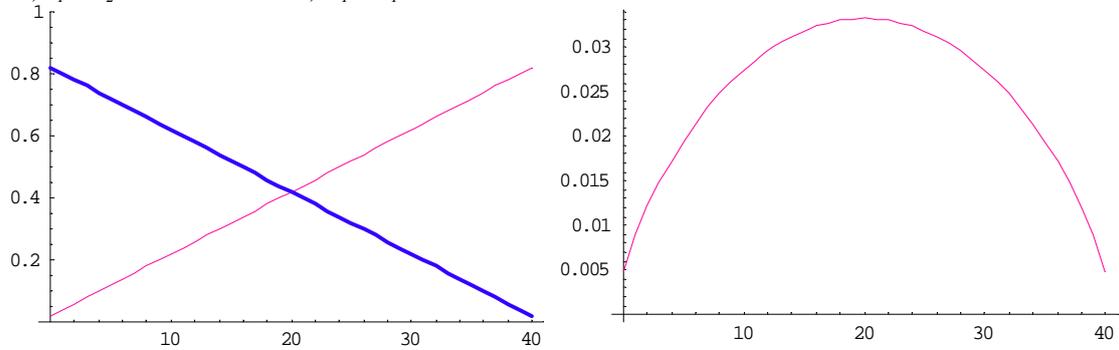
Fig 6. Chen's model. Left $p_+(n)$, $p_-(n)$, right probability density, $n$ is plotted on the x-axis. $N=40$, $a_1 = a_2 = 0.02$; $b_1 = b_2 = 0.8/40 = 0.02$; $b_1 = a_1$

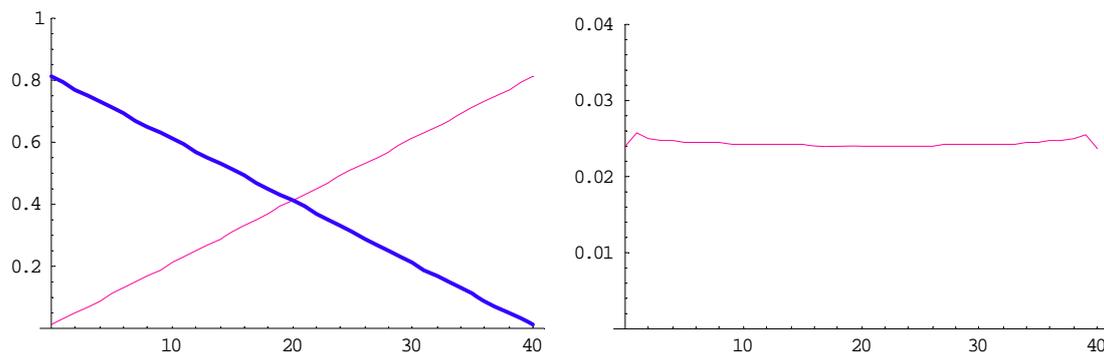

Fig 7. Chen's model. Left $p_+(n)$, $p_-(n)$, right probability density, $n$ is plotted on the x-axis. $N=40$, $a_1 = a_2 = 0.0115$; $b_1 = b_2 = 0.8/40 = 0.02$, $b_1 > a_1$.

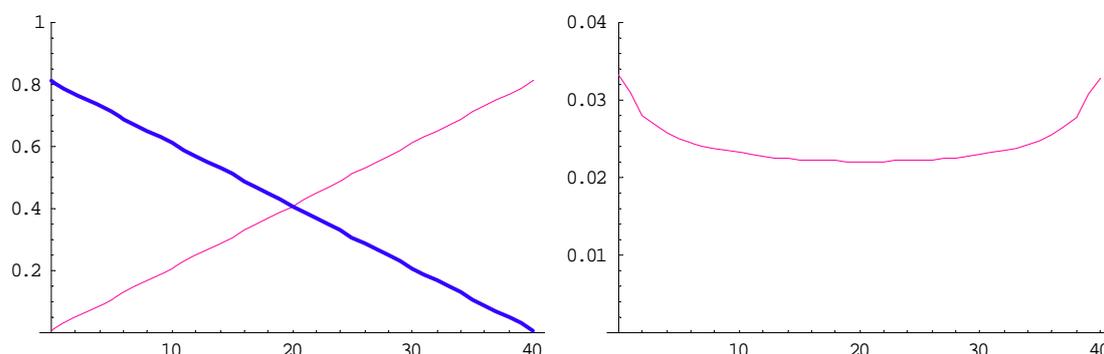

Fig 8. Chen's model. Left $p_+(n)$, $p_-(n)$, right probability density, $n$ is plotted on the x-axis. $N=40$, $a_1 = a_2 = 0.01$; $b_1 = b_2 = 0.8/40 = 0.02$, $b_1 > a_1$.

## *Regime switching (RS) model*

$$p_+(n) = \begin{cases} a_1 \text{ if } n < N_1 \\ b_1 \text{ if } n \geq N_1 \end{cases} \qquad p_-(n) = \begin{cases} a_2 \text{ if } N - n < N_2 \\ b_2 \text{ if } N - n \geq N_2 \end{cases}$$

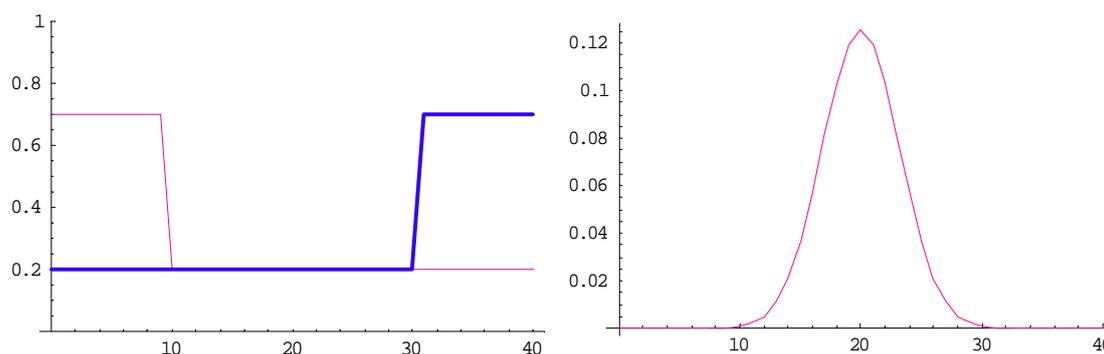

Fig 9. RS Model. Left $p_+(n)$, $p_-(n)$, right probability density, $n$ is plotted on the x-axis. $N=40$, $a_1 = a_2 = 0.2$; $b_1 = b_2 = 0.7$; $N_1 = N_2 = 30$.

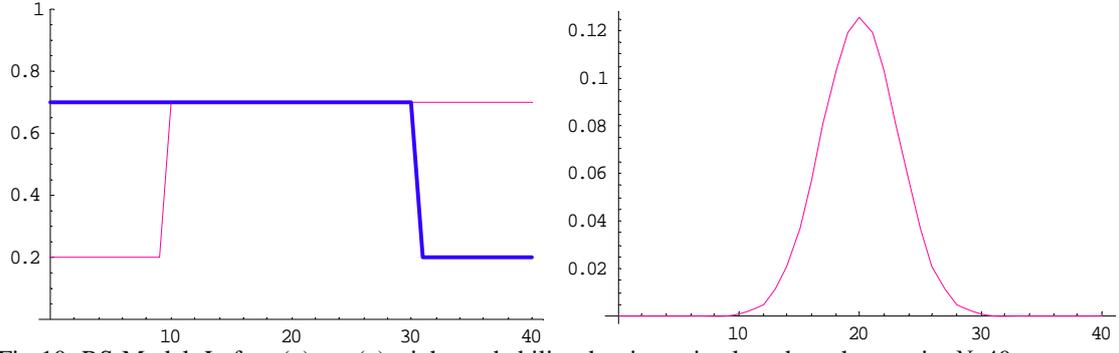

Fig 10. RS Model. Left $p_+(n)$, $p_-(n)$, right probability density, $n$ is plotted on the x-axis. $N=40$, $a_1 = a_2 = 0.7$; $b_1 = b_2 = 0.2$; $N_1 = N_2 = 30$.

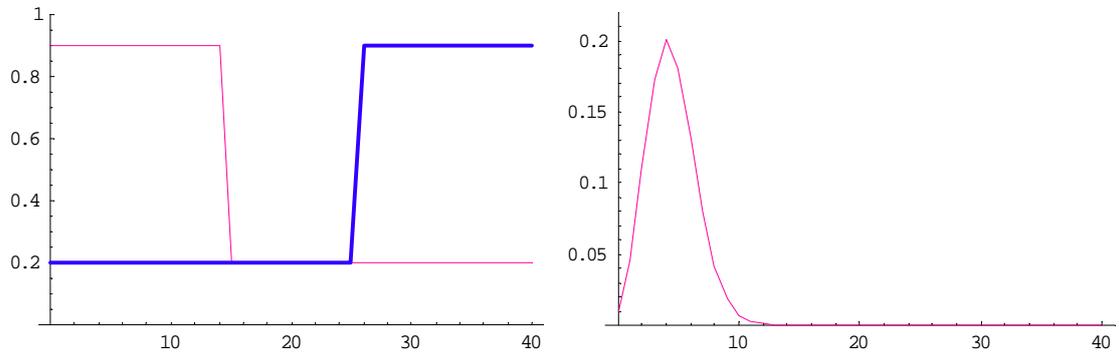

Fig 11. RS Model. Left $p_+(n)$, $p_-(n)$, right probability density, $n$ is plotted on the x-axis. $N=40$, $a_1 = a_2 = 0.2$; $b_1 = b_2 = 0.9$; $N_1 = N_2 = 25$.

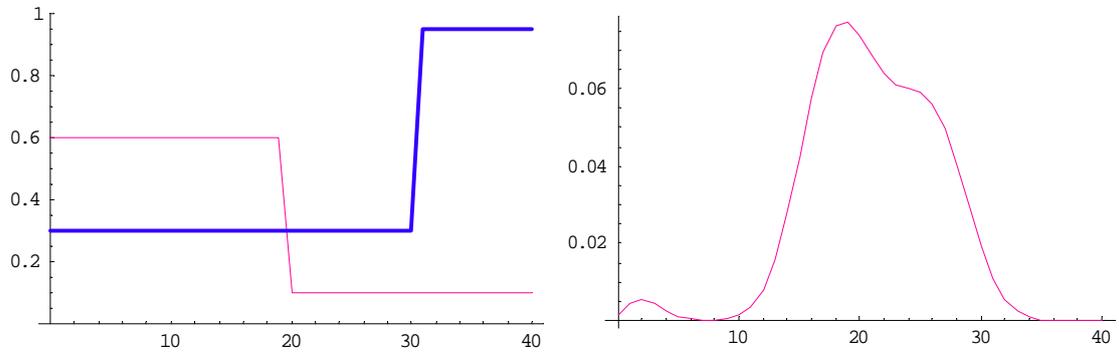

Fig 12. RS Model. Left $p_+(n)$, $p_-(n)$, right probability density, $n$ is plotted on the x-axis. $N=40$, $a_1 = 0.3$, $a_2 = 0.1$; $b_1 = 0.95$, $b_2 = 0.6$; $N_1 = 30$, $N_2 = 20$.

## 3. Types of Markets and Transition States in Social Imitation Model

We consider again the Social Imitation model

$$p_+(n) = c \exp\left(-\left(k\left(\tfrac{n}{N} - \tfrac{1}{2}\right) + h\right)\right),$$
$$p_-(n) = c \exp\left(k\left(\tfrac{n}{N} - \tfrac{1}{2}\right) + h\right),$$

$k$ is any non–negative number, $h$ is any real number.

The theory assumes that the rate of change of the opinion of an individual is enhanced by the group of individuals with the opposite opinion and diminished by people of one's own opinion. The theory assumes also that there exists a sort of overall social climate, which facilitates or hinders the change of opinion. We introduce $x = n/N - 1/2$, and in analogy to the physics Ising model, we may write

$$p_+(x) = c\exp\left(-\frac{Ix+H}{T}\right) = c\exp(-(kx+h)),$$

$$p_-(x) = c\exp\left(\frac{Ix+H}{T}\right) = c\exp(kx+h).$$

Here $I$ is a measure of the strength of adaptation to neighbors, $H$ is a preference parameter ($H > 0$ means that bullish opinion is preferred to bearish), $T$ is a collective climate parameter corresponding to temperature in physics. Numerical solution shows that, depending on the fundamentals and investors' sentiment parameter, probability distribution function may have a number of qualitatively different forms (shown on the figure). Different types of markets exhibited are discussed below.

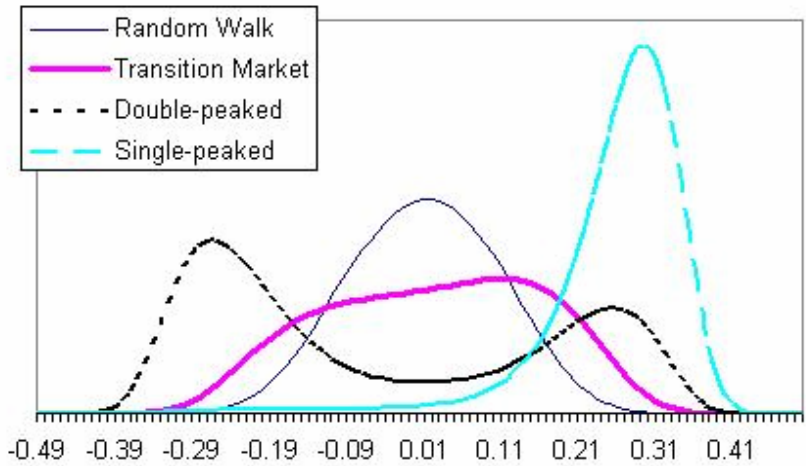

Fig. 13. Types of stationary distributions in Social Imitation model
Horizontal axis is $x = n/N - 1/2$

Market returns may either fluctuate randomly around zero or, under the specific conditions, may exhibit a high degree of polarization, leading to a large net difference between gainers and losers and corresponding large market moves. Ising model described above, when applied to the stock markets, has three key parameters: the sentiment $k$ is a measure of whether the level of coherent behavior is below or above a critical transition threshold. The fundamental bias $h$ is a measure of external preference towards a bullish or bearish opinion. The third parameter $N$ is the number of degrees of freedom of the market, i.e. the typical number of investors participating. Market behavior observed remains qualitatively the same for a sufficiently large number of participants. Below the critical transition threshold, the random walk state prevails. The random-walk market is much less sensitive to changes in fundamentals than it is in the periods of crowd behavior. Above the transition threshold, if fundamentals are highly positive, trending bull markets occur; volatile "double-peaked" markets are expected if fundamentals do not provide a clear direction for investors.

In the case when investors' opinion is not conducive to crowd behavior, the nonlinear model discussed may be considerably simplified. If we assume that fundamentals are zero the probability distribution becomes normal with variance. Thus, market is in the normal random walk state then. Theoretically, random-walk markets could provide either a small, stable bullish return or small and stable bearish return, depending on the fundamentals. Historically, however, random-walk markets have provided a stable negative return and are most frequently associated with bear markets. When nonlinear effects are significant, transition probabilities of increasing and decreasing the share of bullish investors are no longer equal. In effect, steps in one direction may be larger and more likely then steps in the other direction. While the nonlinear model reduces to the usual random walk as a special case, the general nonlinear, time-dependent situation is quite complex.

However, when investors' sentiment grows, crowd behavior becomes dominating and return distribution deviates far from normal. In the next example, market gradually changes its state from random walk to a strongly trending bullish market. If, on the other hand, fundamentals exhibit a drift while the level of crowd behavior does not change, the outcome is very different. In the next figure the case is described when, due to a large change in fundamentals, polarization of investors' crowd eventually changes and leads to an abrupt transition from trending bearish to bullish behavior.

As $k$ approaches the critical transition threshold, the variance of the normal probability distribution in the random walk state grows up and normal distribution no longer applies. The random-walk model is no longer valid during the transition to trending behavior. Instability occurs at transition. This implies a highly inefficient market in which large, long-lasting sentiment oscillations must be expected. Even a slight fundamental bias may skew the distribution strongly.

As $k$ increases above the critical threshold, the model predicts a bimodal probability distribution. That is, a high degree of polarization exists among investors, but without a strong fundamental bias, there is no clear indication of which direction the crowd chooses. Furthermore, there is a possibility of abrupt opinion fluctuations from bullish to bearish and in the opposite direction. The probability of a large shift is greatest when prevailing investor sentiment is in the opposite direction to a small external bias in fundamentals. This is well illustrated by the example of 1987 crash discussed in more detail below. A "double-peaked" market state may be described as quasi-efficient. The market path itself is unpredictable; the high standard deviation of the returns' distribution reflects the high degree of risk in the "double-peaked" market state.

When fundamentals are strongly positive during a period of crowd behavior, a trending bull market may occur. A trending bull market is a market in which the bearish lobe of the probability distribution is very small. The peak of the distribution in this case is even higher than the expected return. However, the distribution has a long tail that goes into the negative semi axis. Thus, even under extremely bullish conditions, there is a small but finite probability that the market will provide a negative return. Most of the market's long-term gains are achieved during trending markets. However, after a trending market period ends, and a double-peaked or random walk state begins, it is too late to invest already.

Trending bear markets are directly opposite to trending bull markets; only in this case crowd behavior with strong bearish fundamentals is required. Now the positive-side lobe of the distribution begins to disappear. The standard deviation range is the same as for the bear trending market; however expected return is now deep in the negative semi axis. Fortunately, in practice trending bear markets are quite rare.

# 4. Simulation and Empirical Results

On its path, the market changes its behavior moving through a series of stable states described above. Different time-dependent market scenarios were constructed using corresponding simulations

In the case when investors' opinion is not conducive to crowd behavior, i.e. when $k$ is below the transition threshold, market is efficient, and its return distribution changes in time according to classical results; it follows arithmetic Brownian motion. However, when investors' sentiment grows, crowd behavior becomes dominating and return distribution deviates far from normal.

If, on the other hand, fundamentals exhibit a drift while the level of crowd behavior does not change, the outcome is very different. In the next figure the case is described when, due to a large change in fundamentals, polarization of investors' crowd eventually changes and leads to an abrupt transition from trending bearish to bullish behavior.

In the case of Social imitation model, there exist only three time-independent stable market states. As argued in section 2, every initial sentiment distribution different from these 3 stable states converges with time to one of those, depending on the values of fundamental parameters.

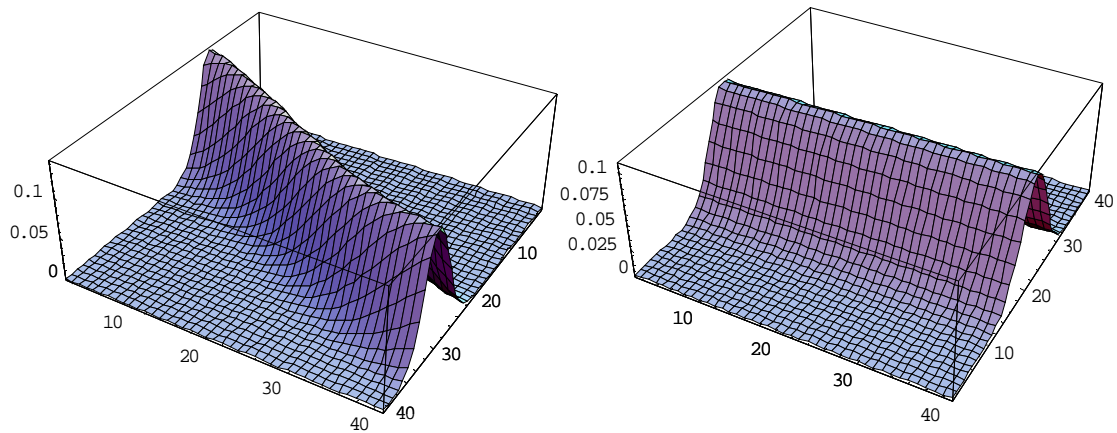

Fig 14. Left: matrix elements of transition matrix $W$, $N=40$; $k=1$; $h=0$; $c=0.45$. One can see that away from diagonal matrix elements are very small. Right: matrix elements of $W^{30}$. Horizontal axes are enumerating $n$ indexes (number of positively oriented investors) in the transition matrix.

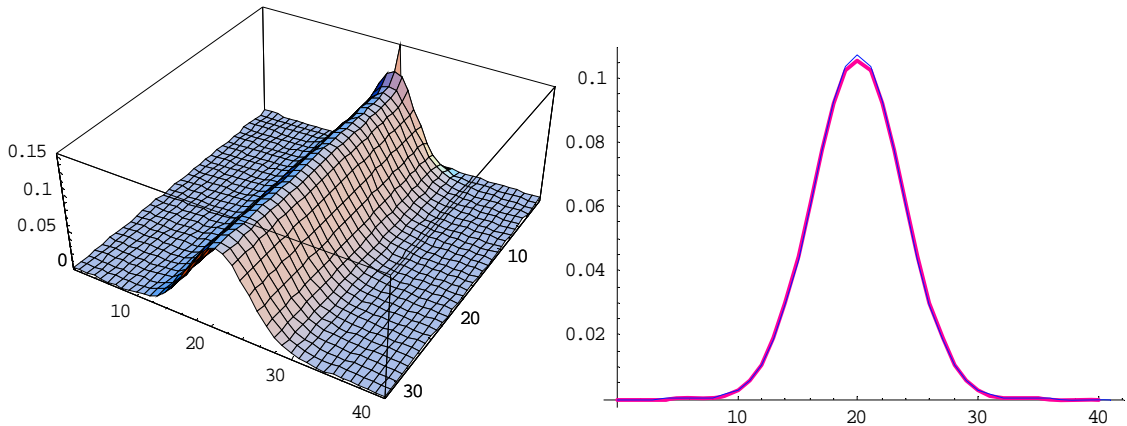

Fig 15. Left: Probability density *f(n, t)*, *t*=0..30 with initial condition *f(n, 0)*= δ(*n* −20), *N*=40; *k*=1; *h*=0; *c*=0.45. Right: density *f(n, 30)* and density of normal distribution with same mean and variance.

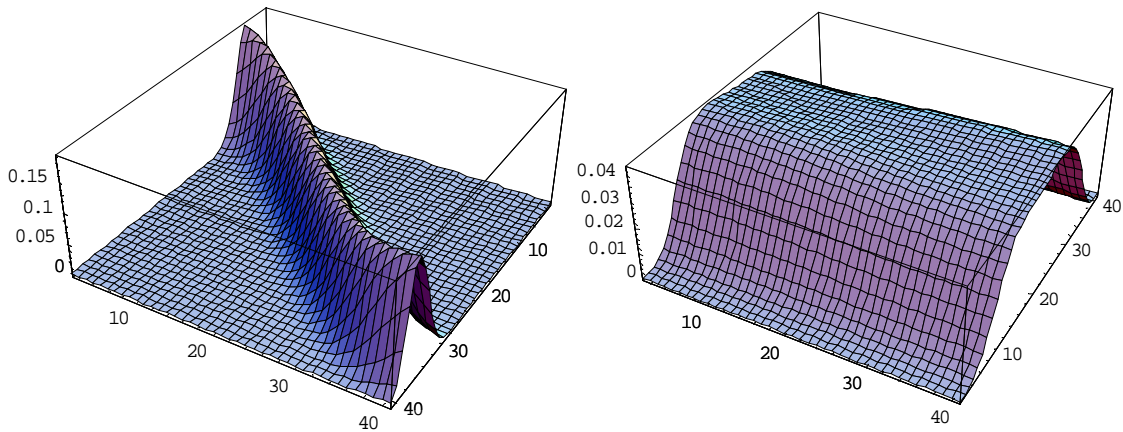

Fig 16. Left: matrix elements of transition matrix *W*, *N*=40; *k*=2; *h*=0; *c*=0.35. One can see that away from diagonal matrix elements are very small. Right: matrix elements of $W^{50}$. Horizontal axes are enumerating *n* indexes (number of positively oriented investors) in the transition matrix.

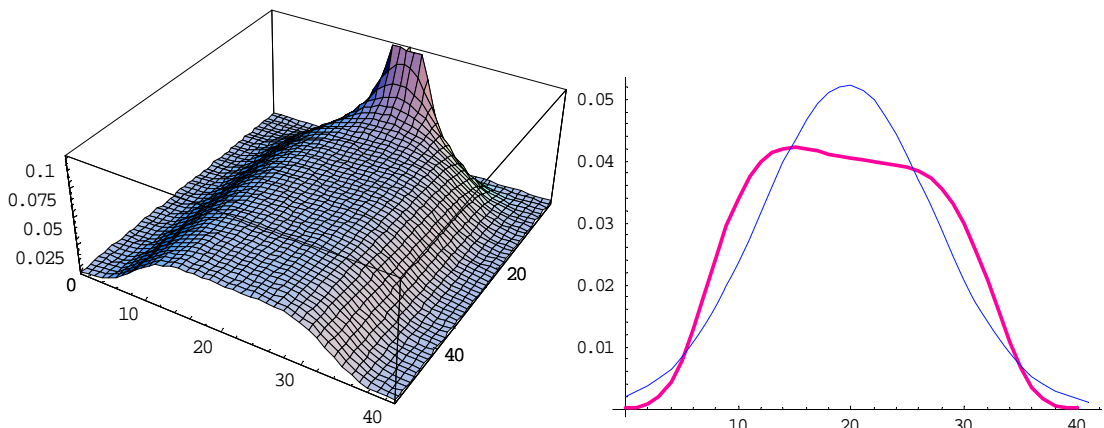

Fig 17. Left: Probability density *f(n, t)*, *t*=0..50 with initial condition *f(n, 0)*= δ(*n* −20), *N*=40; *k*=2; *h*=0; *c*=0.35. Right: density *f(n, 50)* and density of normal distribution with same mean and variance.

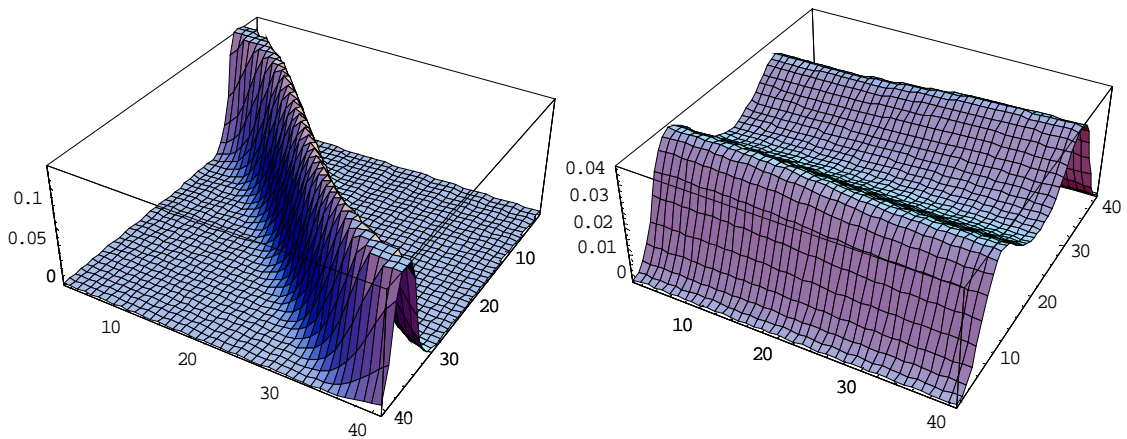

Fig 18. Left: matrix elements of transition matrix $W$, $N=40$; $k=2.2$; $h=0$; $c=0.32$. One can see that away from diagonal matrix elements are very small. Right: matrix elements of $W^{100}$. Horizontal axes are enumerating $n$ indexes (number of positively oriented investors) in the transition matrix.

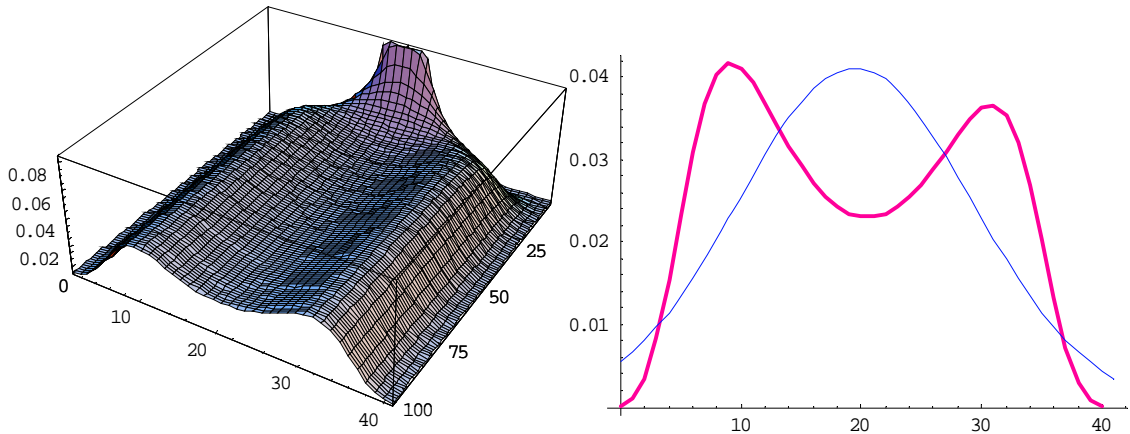

Fig 19. Left: Probability density $f(n, t)$, $t=0..100$ with initial condition $f(n, 0)= \delta(n -20)$, $N=40$; $k=2.2$ $h=0$; $c=0.32$. Right: density $f(n, 100)$ and density of normal distribution with same mean and variance.

A dramatic example of market following a series of states discussed is the market crash in 1987. The climate prior to the crash was conducive to crowd behavior. This was indicated by the extreme in market volume. Fundamentals were small bearish, while the market has risen to an enormously high level of 25% annual return so that bullish sentiments prevailed. Thus, the market was in a less probable state. This eventually led to a large swing in crowd sentiment, causing the Crash. The historical data of the relevant dates are described by the illustrations.

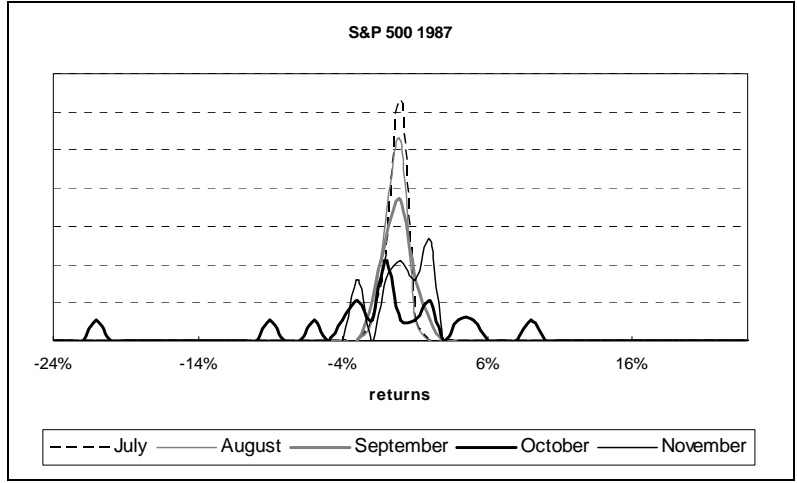

Fig 20. Annualized return distributions of S&P 500 index for several consecutive months in 1987.

The model appears to describe very well market behavior in the vicinity of dramatic events like market crashes. Thus, another good application of this theory is the Japanese stock market crash in 1988. The return distributions in different market states during the year 1988 are described by the figure 21.

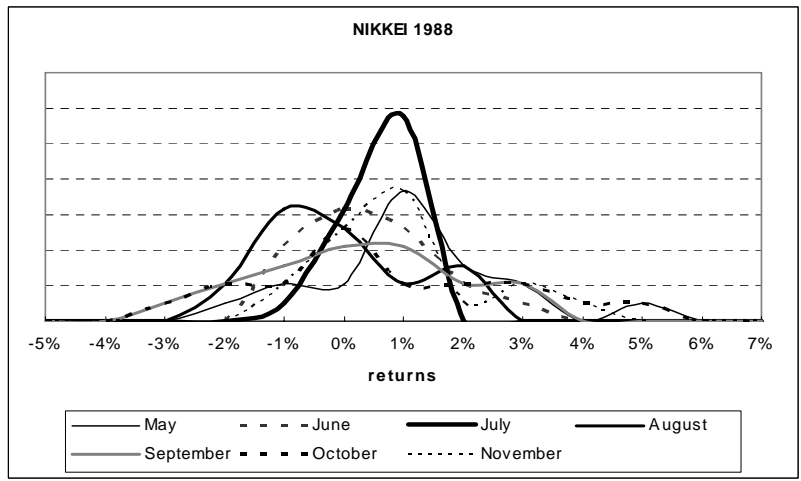

Fig 21. Annualized return distributions of NIKKEI index for several consecutive months in 1988.

The model described here predicts that the long-term up-trend in stock prices is in fact comprised of periods of high reward and low risk mixed with other periods of high risk and low (or negative) returns. This implies that there is less noise in the return distribution than Efficient Market Theory would predict.

There are also many other examples of application of this model to "regular" situations; for instance, we have explored the behavior of several major stock indices during the year 2003. During most of the time the market is in the random walk state or exhibits "double-peaked" behavior without strong fundamentals. Returns' actual distribution does not deviate much from normal, in fact, as predicted by the model.

The underlying assumptions of the model are very general; thus, the hypothesis can be applied to markets other from stock market. Figure 22 exhibits the actual return distributions for Australian Dollar during the year of 2003.

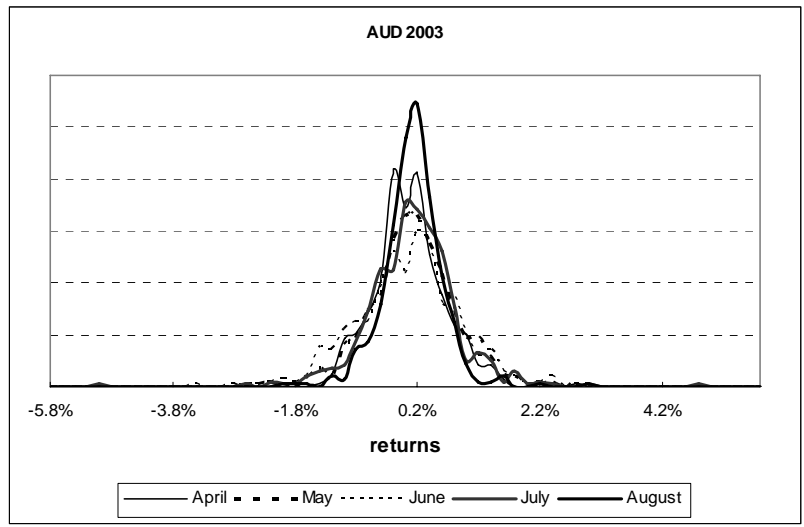

Fig 22. Annualized return distributions of Australian Dollar spot rate for consecutive months in 2003.

We have also considered a futures example, which is a study of generic back-adjusted contracts on different assets. Here we consider not only equity but most major asset classes tradable through futures: major currencies, fixed-income (long- and short-term, U.S. and international), metals, and equity indices. Empirical evidence shows that the result is indeed applicable to other markets such as different derivatives' markets as well. Probability distributions for several futures types are provided at the figure below.

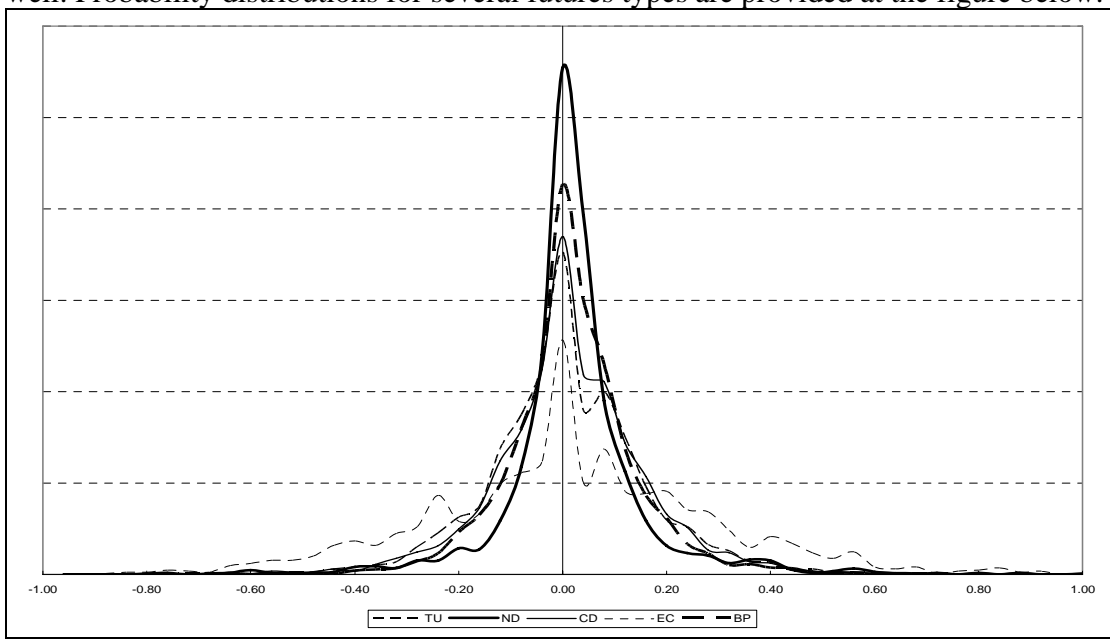

Fig 23. Probability distribution function of returns of different classes of futures contracts in 2004.

Most hedging strategies are designed to trade upside potential off against reduced overall portfolio volatility. The model suggests that it should be possible to both beat the market and reduce risk. During times when market is not in the random walk state the distribution of returns on stocks differs significantly from normal. Thus, probability density functions obtained in the simulation could be used for evaluating portfolio risk, option pricing, etc. instead of normal distribution, and it seems appealing to implement it in practice.

Columbia University, 2990 Broadway, New York, NY 10027, USA